\documentstyle[12pt]{article}
\begin{document}
{\pagestyle{empty}
\vskip 2.5cm
~\\

{\renewcommand{\thefootnote}{\fnsymbol{footnote}}
\centerline{\large \bf New Generalized-Ensemble Monte Carlo Method}
\centerline{\large \bf for Systems with Rough Energy Landscape}
}
\vskip 3.0cm

\centerline{Ulrich H.E.~Hansmann
\footnote{\ \ e-mail: hansmann@ims.ac.jp}
 and Yuko Okamoto
\footnote{\ \ e-mail: okamotoy@ims.ac.jp}}
\vskip 1.5cm
\centerline {{\it Department of Theoretical Studies}}
\centerline {{\it Institute for Molecular Science}}
\centerline {{\it Okazaki, Aichi 444, Japan}}

\medbreak
\vskip 3.5cm
 
\centerline{\bf ABSTRACT}
\vskip 0.3cm
We present a novel Monte Carlo algorithm which enhances 
equilibrization of low-temperature
simulations and allows sampling of configurations over 
a large range of energies. The method is based on a non-Boltzmann
probability weight factor and is another version of the so-called
generalized-ensemble techniques.
The effectiveness of the new approach is demonstrated for
the system of a small peptide, 
an example of the frustrated system with a rugged energy landscape.
\vskip 1.5cm
 
\vfill

\vskip 3.5cm

\newpage}
\baselineskip=0.8cm
\noindent
The energy landscape of
 many  important physical systems is characterized
by a huge number of local minima separated by high energy barriers.
In the canonical ensemble with temperature $T$, the  probability to 
cross an 
energy barrier of heights $\Delta E$ is proportional 
to $e^{-\Delta E/k_B T}$, where $k_B$ is the Boltzmann constant.
Hence, at low temperatures, canonical molecular dynamics and Monte Carlo
simulations 
will get trapped in configurations corresponding to 
one of these local minima.
Only  small parts of the entire phase space can be explored, 
rendering the calculation of physical quantities unreliable.

In principle, one can think of two ways to overcome this
difficulty.
One way is to look for improved updates of configurations
in the numerical simulation.
The cluster algorithm \cite{CA} is an example of global updates 
that enhance thermalization and has
been very successful in spin systems. 
However, for most other systems with frustration,
no such updates are known. 
Another 
way to overcome the supercritical slowing down is to perform a
 simulation in a so-called
{\it generalized ensemble}, which is based on a non-Boltzmann
probability distribution.
Multicanonical algorithm \cite{MU,MU3}, $1/k$-sampling \cite{HS},
 and simulated tempering \cite{L,MP} are prominent  
examples of such an approach.
Common to the three techniques is that a 
molecular dynamics or Monte Carlo
simulation is performed in an artificial ensemble defined in such a way
that a uniform (non-canonical) distribution of the chosen physical 
quantity is 
obtained.  For instance, in the multicanonical algorithm the weight 
$w_{mu}(E)$ 
is chosen so that the distribution of energy is uniform:
\begin{equation}
P(E) \propto n(E)~ w_{mu}(E) = {\rm const},
\label{eq1}
\end{equation}
where $n(E)$
is the density of states.  A simulation based on this weight factor
results in 
a free random walk in the energy 
space.  Hence, the simulation can  
escape from any energy barrier, and 
even regions
with small $n(E)$ can be explored in detail. 
Similarly, $1/k$-sampling yields
a uniform distribution in (microcanonical) entropy, and simulated 
tempering a uniform distribution in temperature. 
The great advantage of these generalized-ensemble methods
lies in the fact that from a single simulation run
one can not only locate the energy global minimum 
but also obtain the canonical distribution for a wide
temperature range by
the reweighting techniques \cite{FS}. 

Despite their successful applications to systems with 
first-order phase 
transitions \cite{MU}, spin glasses \cite{BC}, and the protein 
folding problem \cite{HO,HSC}, 
generalized-ensemble methods are not without problems. Unlike in the canonical
ensemble, the probability weights are not {\it a priori} known.
For instance, for the case of multicanonical algorithm, 
Eq.~(\ref{eq1}) implies
\begin{equation}
w_{mu} (E) \propto n^{-1} (E)~,
\label{eq2}
\end{equation}
and the knowledge of the exact weight would be equivalent 
to obtaining the density of states $n(E)$, i.e., solving the system. 
Hence, one needs its estimator for a numerical simulation. 
The determination of the weight $w_{mu}(E)$ is usually based on
an iterative
procedure first described in Ref.~\cite{MU3}, and can be non-trivial
and tedious. 
In this Letter, we present a new generalized-ensemble algorithm
in which the determination of the weight is simple and
straightforward.

Our aim is to develope a new generalized-ensemble algorithm
in which the determination of the probability weight factor
is simpler.
For this, we try to slightly modify the Boltzmann weight, whereas
other generalized-ensemble approaches use drastically different
weights.
The weight should enhance the thermalization of
low-temperature simulations and ensure sufficient sampling in
the low-energy region. Hence, we are interested in an
ensemble where not only the low-energy region can be sampled 
efficiently but also
the high-energy states can be visited with finite probability. 
The latter 
feature ensures that energy barriers can be overcome and that the 
simulation can escape from
local minima. The probability distribution of energy should resemble that
of an ideal low-temperature Boltzmann distribution,
but with a tail to higher
energies. One choice is that the sampling of low-energy states 
is described
by an exponential function (Boltzmann weight), 
while that of high-energy states follows a
power law. 
Guided by these considerations, we propose the
following as the new weight:
\begin{equation}
w(E) = \left( 1+\beta \frac{E-E_{GS}}{m} \right)^{-m}~,
\label{eqwe}
\end{equation}
where $\beta \equiv \frac{1}{k_B T}$, $E_{GS}$ is
the global-minimum energy,
and $m \ ( > 0 )$ is a free parameter.
Here, we are shifting the zero of energy by $E_{GS}$ in order
to assure that energy is always non-negative.
We remark that weights with the same mathematical structure
also appear in the framework of Tsallis generalized
statistical mechanics \cite{Tsa}, which was developed for
simulations of non-extensive systems (e.g., fractal 
random walks).  An application to optimization problems
can be found in Ref.~\cite{TSA}.

Obviously, the new weight in Eq.~(\ref{eqwe}) reduces to
the canonical Boltzmann weight in the low-energy 
(and hence low-temperature) region
for $\frac{\beta (E-E_{GS})}{m} \ll 1$.
On the other hand, this weight at high energies is no
longer exponentially suppressed, but only according to 
a power law with
the exponent $m$.
Note that our choice of sign in Eq. (\ref{eqwe}) is important. 
>From a mathematical point of view, 
$(1 - \beta \frac{E-E_{GS}}{m})^{m}$ is 
equally a good approximation to the canonical weight,
but is not useful as a weight in numerical simulations,
since the expression inside the parentheses
can become negative. 

In this work we consider a system with continuous degrees of freedom.
At low temperatures the harmonic approximation holds, 
and the density of states is given by
\begin{equation}
n (E) \propto (E - E_{GS})^{\frac{n_F}{2}}~,
\label{eqds}
\end{equation}
where $n_F$ is the number of degrees of freedom of the system
under consideration.
Hence, by Eqs.~(\ref{eqwe}) and (\ref{eqds}) 
the probability distribution of energy for the
present ensemble is given by
\begin{equation}
P(E) \propto n(E) w(E) \propto (E - E_{GS})^{\frac{n_F}{2} - m}~,
\label{eqpd}
\end{equation}
for $\beta \frac{E-E_{GS}}{m} \gg 1$.  This implies that
we need
\begin{equation}
m > \frac{n_F}{2} ~.
\label{eqmlim}
\end{equation}
For, otherwise, the sampling of high-energy configurations will be 
enhanced too much. On the other hand, in the limit $m \rightarrow \infty$
our weight tends for all energies to the Boltzmann weight and high-energy
configurations will not be sampled. 

In order for low-temperature simulations to be able to
escape from energy local minima, the weight should 
start deviating from the (exponentially damped)
Boltzmann weight at the energy near its mean value (because
at low temperatures there are  
only small fluctuations of energy around its mean).
In Eq.~(\ref{eqwe}) we may thus set
\begin{equation}
\beta \frac{<E>_T - E_{GS}}{m} = \frac{1}{2}~.
\label{eqnf}
\end{equation}
The mean value at low temperatures is given by the 
harmonic approximation:
\begin{equation} 
<E>_T  ~= E_{GS} + \frac{n_F}{2} k_B T = E_{GS} + \frac{n_F}{2 \beta}~.
\label{eqhm}
\end{equation}
Substituting this value into Eq.~(\ref{eqnf}), we obtain 
the following optimal value for the exponent $m$:
\begin{equation}
m_{opt} = n_F ~.
\label{eqoptm}
\end{equation}
Hence, the optimal weight factor is given by
\begin{equation}
w(E) = \left( 1+\beta \frac{E-E_0}{n_F} \right)^{-n_F}~,
\label{eqopwe}
\end{equation}
where $E_0$ is the best estimate of the global-minimum energy
$E_{GS}$.
     
We have tested our new method in the system for the
protein folding problem, a long-standing problem in
biophysics with 
rough energy landscape. Here, Met-enkephalin has become an often-used 
model to examine the performance of new algorithms, and we study
the same system. 
Met-enkephalin has the amino-acid sequence Tyr-Gly-Gly-Phe-Met.
The energy function
$E_{tot}$ (in kcal/mol) that we used is given by the sum of
the electrostatic term $E_{C}$, 12-6 Lennard-Jones term $E_{LJ}$, and
hydrogen-bond term $E_{HB}$ for all pairs of atoms in the 
peptide together with
the torsion term $E_{tor}$ for all torsion angles:
\begin{eqnarray}
E_{tot} & = & E_{C} + E_{LJ} + E_{HB} + E_{tor}~,\\
E_{C}  & = & \sum_{(i,j)} \frac{332 q_i q_j}{\epsilon r_{ij}}~,\\
E_{LJ} & = & \sum_{(i,j)} \left( \frac{A_{ij}}{r^{12}_{ij}}
                                - \frac{B_{ij}}{r^6_{ij}} \right),\\
E_{HB}  & = & \sum_{(i,j)} \left( \frac{C_{ij}}{r^{12}_{ij}}
                                - \frac{D_{ij}}{r^{10}_{ij}} \right),\\
E_{tor}& = & \sum_l U_l \left( 1 \pm \cos (n_l \chi_l ) \right).
\end{eqnarray}
Here, $r_{ij}$ is the distance (in {\AA}) between the 
atoms $i$ and $j$, and 
$\chi_l$ is 
the torsion angle for the chemical bond $l$. 
The parameters for the energy function and the molecular
geometry (with fixed bond lengths and bond angles) were adopted from
ECEPP/2 (Empirical Conformational Energy Program for Peptides) 
\cite{EC3}.  
The dielectric constant $\epsilon$ was set equal to 2.
Fixing the peptide bond angles $\omega$ to 
$180^{\circ}$ leaves us with 19 torsion angles as independent
degrees of freedom (i.e., $n_F = 19$).
The computer code KONF90 \cite{KONF}  was
used.  One Monte Carlo sweep updates every torsion angle 
of the peptide once.

It is known from our previous work that the global-minimum value
of KONF90 energy for Met-enkephalin is
$E_{GS} = -12.2$ kcal/mol \cite{HO94_3}.  The peptide has essentially
a unique three-dimensional structure at temperatures
$T \le 50$ K, and the
average energy is about $-11$ kcal/mol at $T = 50$ K \cite{HO}.
Hence, in the present work we always set $T= 50$ K (or, $\beta = 10.1$
$[\frac{1}{{\rm kcal}/{\rm mol}}]$) in our new probability
weight factor.
All simulations were started from completely random initial
configurations (Hot Start).

To demonstrate that thermalization is greatly enhanced in our ensemble, 
we first compare the \lq \lq time series\rq \rq~ 
of energy as a function of Monte Carlo
sweep.  In Fig.~1 we show the results 
from a regular canonical Monte Carlo simulation
at temperature $T = 50$ K (dotted curve) and 
those from a gereralized-ensemble 
simulation of the new algorithm (solid curve).
Here, the weight we used for the latter simulation is given
by Eq.~(\ref{eqopwe}) with $n_F=19$  and 
$E_0 = E_{GS} = -12.2$ kcal/mol. 
For the canonical run the curve stays around the value
$E = -6$ kcal/mol with small thermal fluctuations, reflecting the
low-temperature nature.  The run has apparently been trapped in a local
minimum, since the mean energy at this temperature is
$<E> = -11.1$ kcal/mol as found by a multicanonical simulation
in Ref.~\cite{HO94_3}.
On the other hand, the simulation based on the new weight
covers a much wider energy range than
the canonical run.  It is a random walk in energy
space, which keeps the simulation from getting trapped in a local minimum.
It indeed visits the ground-state region several times
in 200,000 Monte Carlo sweeps.
These properties are common features of generalized-ensemble methods.

Since the simulation by the present algorithm samples a
large range of energies, we can 
use the reweighting techniques \cite{FS} to construct 
canonical distributions and calculate thermodynamic quantities 
over a wide temperature range.
Following 10,000 sweeps for thermalization, we
 performed a single simulation
of 1,000,000 Monte Carlo sweeps,
storing the configuration information at every second sweep.
We have set again $E_0 = -12.2$ kcal/mol and 
$n_F = 19$ in the
weight of Eq.~(\ref{eqopwe}). 
>From this production run one can calculate various thermodynamic quantities
as a function of temperature.  As examples we show the average
energy and the specific heat in Fig.~2a and Fig.~2b, respectively.
The specific
heat here is defined by the following equation:
\begin{equation}
  C \equiv \frac{1}{k_B} \frac{d \left( \frac{<E_{tot}>_T}{N} \right)}{d T} 
= {\beta}^2 \ \frac{<E_{tot}^2>_T - <E_{tot}>_T^2}{N}~, 
\label{eqsh}
\end{equation}
where $N \ (=5)$ is the
number of amino-acid residues in the peptide.  
The harmonic approximation holds at low temperatures,
and by substituting Eq.~(\ref{eqhm}) into Eq.~(\ref{eqsh}), we have
\begin{equation}
  C = \frac{n_F}{2 N} = 1.9~. 
\label{eqshha}
\end{equation}
Note that the curve in Fig.~2b approaches this value in the
$T \rightarrow 0$ limit.
The results from a multicanonical production run with the same
statistics are also shown in the Figures for comparison.  The results from
both methods are in complete agreement.  

We now examine the dependence
of the simulations
on the values of the exponent $m$ in our weight (see
Eqs.~(\ref{eqwe}) and (\ref{eqopwe})) 
and demonstrate that $m = n_F$ is indeed the optimal choice.
Setting $E_0 = E_{GS} = -12.2$ kcal/mol, 
we performed 10 independent simulation runs of  50,000 
Monte Carlo sweeps with 
various choices of $m$.  In Table~I we list the lowest energies
obtained during each of the 10 runs for five choices of
$m$ values: $9.5~(=\frac{n_F}{2})$, 14, $19~(=n_F)$, 50, and 100.  
The results from regular canonical
simulations 
at $T = 50$ K with 50,000 Monte Carlo sweeps are
also listed in the Table for comparison.
If $m$ is chosen to be too small (e.g., $m=9.5$), then
the weight 
follows a power law in which the suppression for higher energy
region is insufficient (see Eq.~(\ref{eqpd})).
As a result, the simulations tend to stay at high energies and
fail to sample 
low-energy configurations. On the other hand, for too large a value 
of $m$ (e.g., $m=100$),
the weight is too close to the canonical weight, and therefore 
the simulations will get trapped in local minima.
It is clear from the Table that $m = n_F$ is the optimal choice.
In this case the simulations found the ground-state configurations
80 \% of the time (8 runs out of 10 runs).
This should be compared with 90 \%, 75 \%, 80 \%, and 40 \%
for multicanonical annealing, $1/k$-annealing, simulated tempering
annealing, and simulated annealing algorithms, respectively,
in simulations with the same number of Monte Carlo sweeps \cite{HO96c}.

To analyze the above results further, we calculated the 
actual probability distributions of energy for various values of
$m$.  This can be done by the reweighting techniques from the
single production run of 1,000,000 Monte Carlo sweeps mentioned above
(which is based on the
weight of Eq.~(\ref{eqopwe}) with $E_0 = -12.2$ kcal/mol 
and $m = n_F = 19$).  The results are shown in Fig.~3a.
By examining the Figure, we again find that $m = n_F$ is the 
optimal choice.  It yields 
to an  energy  distribution which has a pronounced peak
around the mean energy value ($<E>~ = -11.1$ kcal/mol)
at $T=50$ K.
At the same time, it has a tail to higher energies. This behavior 
is exactly what we were looking for and justifies our definition 
of weights in Eq.~(\ref{eqopwe}).

The greatest advantage of the new method over other generalized-ensemble
approaches is the simplicity of the weight factor.
In multicanonical algorithms, $1/k$-sampling, or simulated tempering,
the explicit functional forms of the weights are not known {\it a priori}
and they have to be determined numerically by iterations of
trial simulations.  This can be a formidable task in many cases.
On the other hand,
the  weight factor of the present algorithm just depends on the 
knowledge of the global-minimum energy $E_{GS}$ (see Eq.~(\ref{eqopwe})).
If its value  is known,
which is the case for some systems with frustration, the  weight
is completely determined. However, if $E_{GS}$ is not known, we have to
obtain its best estimate $E_0$.
We can calculate the 
actual probability distributions of energy for various values of
$E_0$ by the reweighting techniques again.  
The results are shown in Fig.~3b.
We see that for the system of Met-enkephalin,
one needs the accuracy of about $1 \sim 2$ kcal/mol in the estimate of
the global-minimum energy $E_{GS}$ in order for our new algorithm
to be effective. This implication is supported by Table~II where we list
the lowest energies obtained during each of 10 independent simulation runs
of 200,000 Monte Carlo sweeps with $m=n_F=19$. Four choices were considered
for the $E_0$ value:  $-12.2,~ -13.2,~ -14.2$, and $-15.2$ kcal/mol. 
We remark that $E_0$
has to underestimate $E_{GS}$ to ensure that $E-E_0$  can not become
negative. Our data show again that an accuracy of $1 \sim 2$ kcal/mol
in the estimate of the global-minimum energy is required for
Met-enkephalin.

The use of our method therefore depends  on  the ability to find
a good estimate for the ground-state energy $E_{GS}$, which is 
still much 
easier than the determination of
the weights for other generalized-ensemble algorithms.
In principle, such estimates
can be found in an iterative way.
Here, we give one of the effective iteration procedures.
One first sets an initial guess of the optimal $E_{0}$ which should
be lower than $E_{GS}$.
One performs a simulation with the weight of the present method
with small number of Monte Carlo sweeps. From this simulation 
one calculates the average energy $<E>_T$ at the chosen 
temperature $T$ by the reweighting techniques. If
 $<E> -~ E_0 \gg \frac{n_F}{2}k_B T$, one raises the value of  
$E_0$ by a certain amount and repeats the short simulation.
One iterates this process until 
 $<E> -~ E_0 \approx  \frac{n_F}{2}k_B T$.  
The search of the optimal $E_0$ can be further 
facilitated by information such as the average energy and 
the specific heat obtained
from high temperature simulations. 
For Met-enkephalin the incorporation of such information gave
a start value of $E_0=-13.8$ kcal/mol, which is already within
the $2$ kcal/mol accuracy required by our method (see Ref.~\cite{uli}
for details).

In summary, we have introduced a new generalized-ensemble
algorithm for simulations of systems with frustration.
We have demonstrated the effectiveness of the method
by taking the example of the system of a small peptide,
Met-enkephalin,
which has a rough energy landscape with a huge number of
local minima.
The advantage of the new method lies in the fact that
the determination of the probability weight factor
is much simpler than in other generalized-ensemble
approaches.

\vspace{0.5cm}
\noindent
{\bf Acknowledgements}: \\
The simulations were performed on the computers at the Computer
Center at the Institute for Molecular Science (IMS), Okazaki,
Japan.  This work is supported, in part, by Grants-in-Aid
for Scientific Research from the Japanese Ministry of
Education, Science, Sports, and Culture.  \\


\newpage
\noindent
{\Large Table Captions:}\\
\begin{enumerate}
\item  Lowest energy (in kcal/mol) obtained by the present
method with several different choices of the free parameter $m$.
The other free parameter $E_0$ was fixed at the value of 
the global-minimum energy $E_{GS} = -12.2$ kcal/mol.
The temperature was set to $T = 50$ K.
The case for $m = \infty$ stands for a regular canonical
run at $T = 50$ K.
For all cases, the total number of Monte Carlo sweeps per run
was 50,000. 
$<E>$ is the average of the lowest energy obtained by the 10 runs
(with the standard deviations in parentheses), 
and $n_{GS}$ is
the number of runs in which a conformation with $E \le -11.0$ kcal/mol
(the average energy at $T=50$ K) was obtained. 
\item  Lowest energy (in kcal/mol) obtained by the present
method with several different choices of the free parameter $E_0$.
The other free parameter $m$ was fixed at the optimal value of 
$n_F = 19$, the number of degrees of freedom.
The temperature was set to $T = 50$ K.
For all cases, the total number of Monte Carlo sweeps per run
was 200,000. 
$<E>$ is the average of the lowest energy obtained by the 10 runs
(with the standard deviations in parentheses), 
and $n_{GS}$ is
the number of runs in which a conformation with $E \le -11.0$ kcal/mol
(the average energy at $T=50$ K) was obtained. 
\end{enumerate}

\newpage
Table~I.\\
\begin{table}[h]
\begin{center}
\begin{tabular}{ccccccc}\hline \hline
\rule{0mm}{1.5mm} \\
$E_0$ & $E_{GS}=-12.2$ & $-12.2$ & $-12.2$ & $-12.2$ & $-12.2$ & \\ 
\rule{0mm}{1.5mm} \\
\hline
\rule{0mm}{1.5mm} \\
$m$ & $\frac{n_F}{2} = 9.5$ & 14 & $n_F = 19$ & 50 & 100 & $\infty$  \\ 
\rule{0mm}{1.5mm} \\
\hline
\rule{0mm}{1.5mm} \\
Run &  &  &  &  &  &  \\
1  & 0.8 & $-5.2$ & $-11.8$ & $-6.9$ & $-6.8$ & $-4.2$ \\
2  & $-1.4$ & $-2.6$ & $-11.5$ & $-7.1$ & $-7.7$ & $-5.2$ \\
3  & 0.1 & $-6.8$ & $-11.5$ & $-6.9$ & $-4.9$ & $-11.8$ \\
4  & 0.5 & $-5.5$ & $-11.7$ & $-8.2$ & $-9.9$ & $-7.1$ \\
5  & $-1.0$ & $-3.4$ & $-11.6$ & $-7.4$ & $-12.0$ & $-3.3$ \\
6  & 1.1 & $-6.4$ & $-11.6$ & $-10.1$ & $-8.8$ & 0.9 \\
7  & $-1.3$ & $-5.1$ & $-8.5$ & $-8.7$ & $-8.7$ & $-5.3$ \\
8  & 0.4 & $-3.3$ & $-9.7$ & $-10.8$ & $-9.5$ & $-6.3$ \\
9  & 1.2 & $-8.1$ & $-11.6$ & $-12.0$ & $-6.8$ & $-6.4$ \\
10 & 1.2 & $-3.3$ & $-11.9$ & $-10.8$ & $-9.5$ & $-4.7$ \\ 
\rule{0mm}{1.5mm} \\
\hline
\rule{0mm}{1.5mm} \\
$<E>$ & $0.2~(1.0)$ & $-5.0~(1.8)$ & $-11.1~(1.1)$ & $-8.9~(1.9)$ & $-8.5~(2.0)$ & $-5.3$~(3.2) \\ 
\rule{0mm}{1.5mm} \\
\hline
\rule{0mm}{1.5mm} \\
$n_{GS}$& 0/10 & 0/10 & 8/10 & 1/10 & 1/10 & 1/10 \\ 
\rule{0mm}{1.5mm} \\
\hline \hline
\end{tabular}
\end{center}
\label{tab1}
\end{table}

\newpage
Table~II.\\
\begin{table}[h]
\begin{center}
\begin{tabular}{ccccc}\hline \hline
\rule{0mm}{1.5mm} \\
$E_0$ & $E_{GS}=-12.2$ & $-13.2$ & $-14.2$ & $-15.2$ \\ 
\rule{0mm}{1.5mm} \\
\hline
\rule{0mm}{1.5mm} \\
$m$ & $n_F = 19$ & 19 & 19 & 19 \\ 
\rule{0mm}{1.5mm} \\
\hline
\rule{0mm}{1.5mm} \\
Run &  &  &  & \\
1  & $-11.8$ & $-11.1$ & $-10.5$ & $-9.0$ \\
2  & $-11.9$ & $-10.8$ & $-8.3$ & $-10.3$ \\
3  & $-11.9$ & $-11.3$ & $-11.6$ & $-9.7$ \\
4  & $-11.9$ & $-10.2$ & $-10.9$ & $-10.8$ \\
5  & $-11.8$ & $-11.2$ & $-6.9$ & $-9.2$ \\
6  & $-11.3$ & $-11.5$ & $-10.8$ & $-9.6$ \\
7  & $-11.9$ & $-11.3$ & $-8.3$ & $-10.3$ \\
8  & $-11.8$ & $-11.4$ & $-5.9$ & $-6.8$ \\
9  & $-12.0$ & $-11.5$ & $-10.6$ & $-8.6$ \\
10 & $-11.7$ & $-10.0$ & $-10.3$ & $-8.9$ \\ 
\rule{0mm}{1.5mm} \\
\hline
\rule{0mm}{1.5mm} \\
$<E>$ & $-11.8~(0.2)$ & $-11.0~(0.5)$ & $-9.4~(1.9)$ & $-9.3~(1.1)$ \\ 
\rule{0mm}{1.5mm} \\
\hline
\rule{0mm}{1.5mm} \\
$n_{GS}$& 10/10 & 7/10 & 1/10 & 0/10 \\ 
\rule{0mm}{1.5mm} \\
\hline \hline
\end{tabular}
\end{center}
\label{tab2}
\end{table}

\newpage
\noindent
{\Large Figure Captions:}\\
\begin{enumerate}
\item  Time series of the total energy $E_{tot}$ (kcal/mol) 
       from a regular canonical simulation 
       at temperature $T = 50$ K (dotted curve) and that
       from a simulation of the present method with the parameters:
       $E_0 = -12.2$ kcal/mol, $m=n_F=19$, and $T=50$ K (solid curve).
\item  Average energy (a) and specific heat (b) as a function 
       of temperature. 
       They were calculated by the reweighting techniques 
       from a single simulation run 
       of the present method with the parameters:
       $E_0 = -12.2$ kcal/mol, $m=n_F=19$, and $T=50$ K.
       The results from a multicanonical simulation are also shown
       for comparison.
        In both simulations (by the present method and by the multicanonical
        algorithm) the total number of Monte Carlo
        sweeps was 1,000,000. 
\item Distributions of energy for various values of the exponent $m$
       (a) and the global-minimum energy estimate $E_0$ (b)
       in the present method.
       The ordinate for (a) is logarithmic.
       The results were obtained by the reweighting techniques 
       from a single simulation run 
       with the parameters:
       $E_0 = -12.2$ kcal/mol, $m=n_F=19$, and $T=50$ K.
       The total number of Monte Carlo sweeps was 1,000,000.
       For (a) the regular canonical distribtion at $T = 50$ K as
       calculated by the reweighting techniques is also shown for
       comparison.
\end{enumerate}
   

\noindent
\begin{thebibliography}{(00)}
\bibitem{CA} R.H. Swendsen and J.-S. Wang, Phys. Rev. Lett.
  {\bf 58}, 86 (1987). 
\bibitem{MU} B.A. Berg and T. Neuhaus, Phys. Lett.
  B {\bf 267}, 249 (1991); Phys. Rev. Lett.
  {\bf 68}, 9 (1992).
\bibitem{MU3} B.A. Berg, Int.~J.~Mod.~Phys. C {\bf 3},
  1083 (1992).
\bibitem{HS} B.~Hesselbo and R.B.~Stinchcombe,\ Phys.~Rev.~Lett.
	     {\bf 74}, 2151 (1995).
\bibitem{L}  A.P.~Lyubartsev,~A.A.Martinovski,\ S.V.~Shevkunov, and \
	     P.N.\ Vorontsov-Velyaminov,\  J.~Chem.~Phys. {\bf 96},
	     1776 (1992).
\bibitem{MP} E.~Marinari and G.~Parisi, Europhys.~Lett. {\bf 19},
	     451 (1992).
\bibitem{FS} A.M. Ferrenberg and R.H. Swendsen, Phys.\ Rev.\ Lett.
  {\bf  61}, 2635 (1988); Phys. Rev. Lett. {\bf 63}, 
1658(E) (1989), and
  references given in the erratum.
\bibitem{BC} B.A. Berg and T. Celik, Phys. Rev. Lett. {\bf 69}, 2292
  (1992); B.A. Berg, T. Celik, and U.H.E. Hansmann,
  Phys. Rev. B {\bf 50}, 16444 (1994).
\bibitem{HO} U.H.E. Hansmann and Y. Okamoto, J.~Comp.~Chem. 
  {\bf 14}, 1333 (1993).
\bibitem{HSC} M.-H. Hao and H.A. Scheraga, J.~Phys.~Chem. 
  {\bf 98}, 4940 (1994).
\bibitem{Tsa}  C.~Tsallis, {\it J.~Stat.~Phys.} {\bf 52}, 479 (1988).
\bibitem{TSA}  D.A.~Stariolo and C.~Tsallis, 
   {\it Annual Reviews of Computational Physics II}, edited by D. Stauffer
   (World Scientific, Singapore, 1995), p. 343;  I. Andricioaei and
   J.E. Straub, {\it Phys. Rev. E} {\bf 53}, R3055 (1996).
J. Phys. Chem. {\bf 88}, 6231~(1984), and references therein.
\bibitem{KONF} H. Kawai, Y. Okamoto, M. Fukugita, T. Nakazawa, and  
  T. Kikuchi, Chem. Lett. {\bf 1991}, 213 (1991);
  Y. Okamoto, M. Fukugita, T. Nakazawa, and 
  H. Kawai, Protein Engineering {\bf 4}, 639 (1991).
\bibitem{HO94_3} U.H.E. Hansmann and Y. Okamoto, J. Phys. Soc. Jpn.
{\bf 63}, 3945 (1994); Physica A {\bf 212}, 415 (1994).
\bibitem{HO96c} U.H.E. Hansmann and Y. Okamoto, {\it J.~Comp.Chem.} {\bf 18}
                920 (1997). 
\bibitem{uli} U.H.E. Hansmann, manuscript in preparation.

\end{thebibliography}
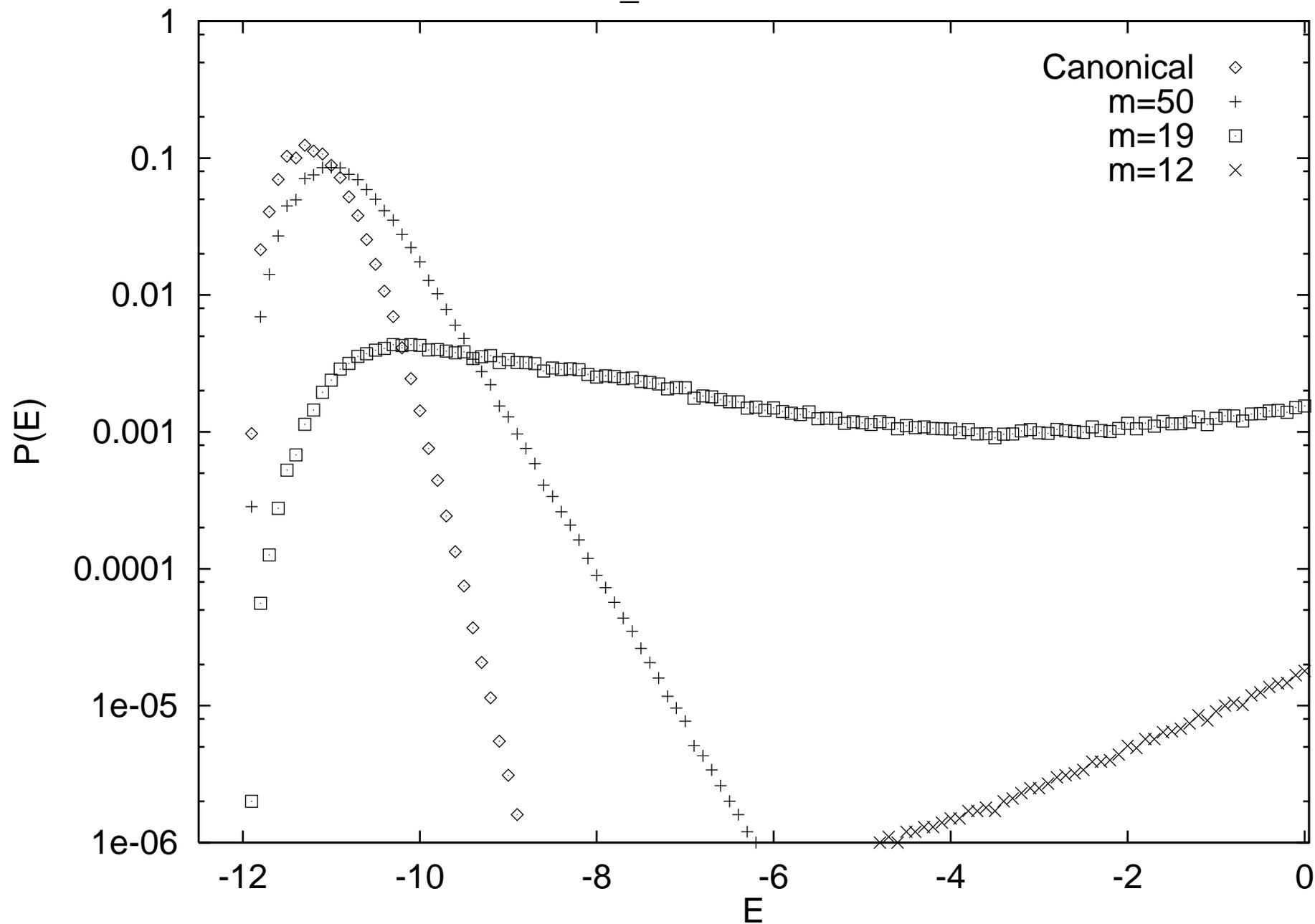
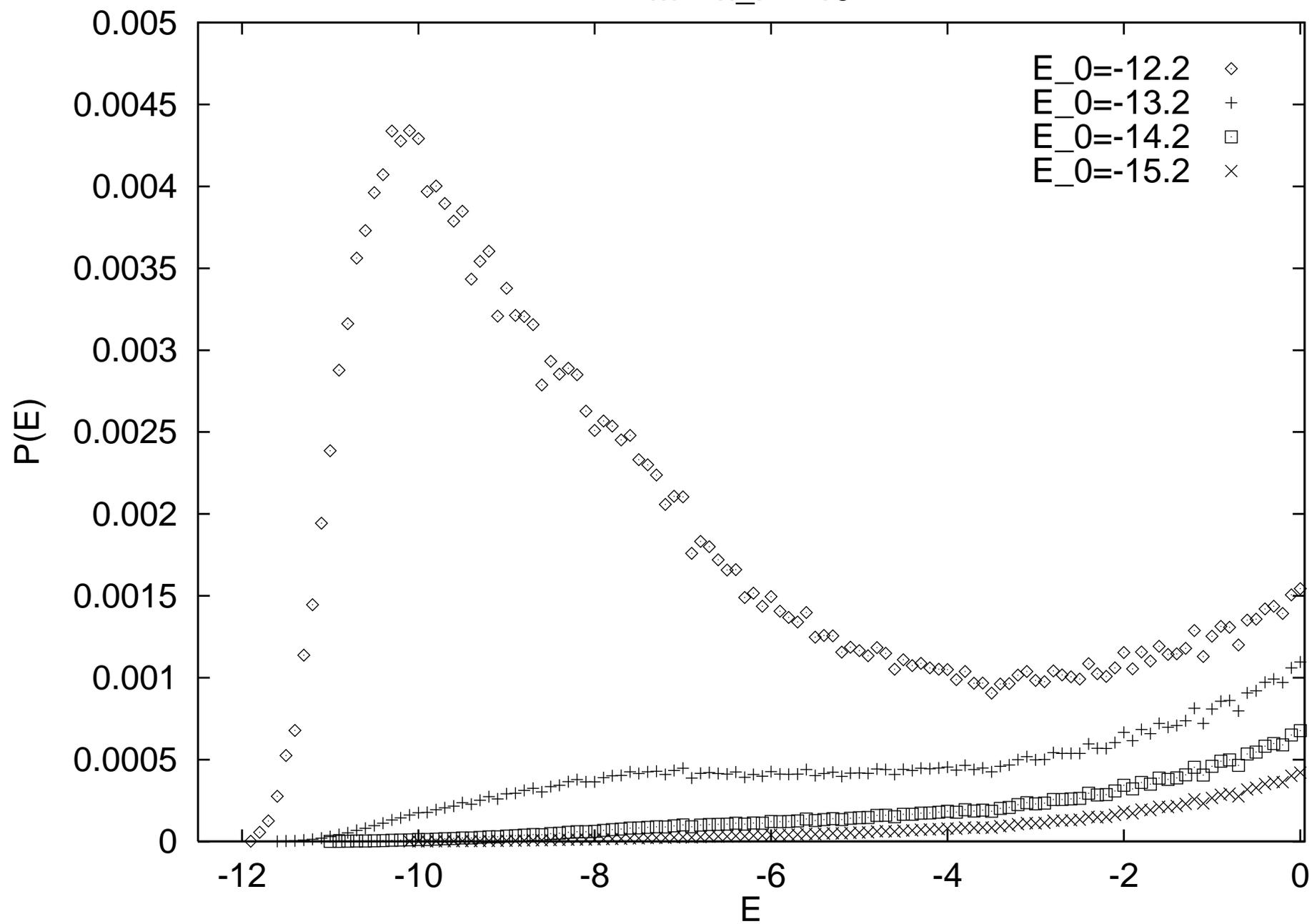
\end{document}